\documentclass[12 pt, amsfonts,amsmath, amssymb,color]{article}
\usepackage{authblk}
\def\baselinestretch{1.0}
\usepackage{amsmath,amssymb,amsfonts,latexsym,float,graphics,epsfig}
\usepackage{subfig}
\usepackage{verbatim}
\usepackage{relsize}
\usepackage{hyperref}
\usepackage{graphicx}
\usepackage{bm}
\usepackage{booktabs}
\usepackage{lmodern}
\usepackage{epstopdf}
\usepackage{slashed}
\usepackage{color}
\usepackage{tikz-cd}
\usepackage[utf8]{inputenc}
\usepackage[T1]{fontenc}

\def\be{\begin{equation}}
\def\ee{\end{equation}}
\def\bea{\begin{eqnarray}}
\def\eea{\end{eqnarray}}

\begin{document}

\renewcommand\theequation{\arabic{section}.\arabic{equation}}
\catcode`@=11 \@addtoreset{equation}{section}

\newtheorem{axiom}{Definition}[section]
\newtheorem{theorem}{Theorem}[section]
\newtheorem{axiom2}{Example}[section]
\newtheorem{lem}{Lemma}[section]
\newtheorem{prop}{Proposition}[section]
\newtheorem{cor}{Corollary}[section]

\newcommand{\ben}{\begin{equation*}}
\newcommand{\een}{\end{equation*}}

\let\endtitlepage\relax

\begin{titlepage}
\begin{center}
\renewcommand{\baselinestretch}{1.5}  

\vspace*{-0.5cm}

{\Large \bf{Quasi-harmonic spectra from branched Hamiltonians}}

\vspace{5mm}
\renewcommand{\baselinestretch}{1}  

\centerline{}

\centerline{{\bf Aritra Ghosh\footnote{{\fontsize{12pt}{14pt}\selectfont \textbf{Present Address:} School of Physics and Astronomy, Rochester Institute of Technology, Rochester, New York 14623, USA}}$^*$, Bijan Bagchi$^\dagger$, A. Ghose-Choudhury$^\ddagger$,}}
\centerline{{\bf Partha Guha$^\#$, Miloslav Znojil$^\$$}}

\vspace{3mm}
\normalsize
\text{$^*$School of Basic Sciences, Indian Institute of Technology Bhubaneswar,}\\
\text{Khurda 752050, Odisha, India}\\
\vspace{1.5mm}
\text{$^\dagger$Department of Applied Mathematics, University of Calcutta,}\\
\text{Kolkata 700009, West Bengal, India}\\
\vspace{1.5mm}
\text{$^\ddagger$Department of Physics, Diamond Harbour Women’s University,}\\
\text{D.H. Road, Sarisha 743368, West Bengal, India}\\
\vspace{1.5mm}
\text{$^\#$Department of Mathematics, Khalifa University of Science and Technology,}\\
\text{Main Campus, P.O. Box - 127788, Abu Dhabi, United Arab Emirates}\\
\vspace{1.5mm}
\text{$^\$$Department of Physics, Faculty of Science, University of Hradec Kr\'{a}lov\'{e},}\\
\text{Rokitansk\'{e}ho 62, 50003 Hradec Kr\'{a}lov\'{e}, Czech Republic}\\

\vspace{7mm}

\text{\textbf{Email:} aritraghosh500@gmail.com}

\vspace{0.4cm}

\begin{abstract}
We revisit the canonical quantization to assess the spectrum of the modified Emden equation $\ddot{x} + kx\dot{x} + \omega^2 x + \frac{k^2}{9}x^3 = 0$, which is an isochronous case of the Li\'enard-Kukles equation. While its classical isochronicity and canonical quantization, leading to polynomial solutions with an exactly-equispaced spectrum have been discussed earlier, including in the recent paper [Int. J. Theor. Phys. 64, 212 (2025)], the present study focuses on the quantization of its branched Hamiltonians. For small $k$, we show numerically that the resulting energy spectrum is no longer perfectly harmonic but only approximately equispaced, exhibiting quasi-harmonic behavior characterized by deviations from uniform spacing. Our numerical results are precisely validated by analytical calculations based on perturbation theory. 
 \end{abstract}
\end{center}
\vspace*{0cm}

\end{titlepage}
\vspace*{0cm}

\clearpage

\section{Introduction}\label{intro_sec}
Oscillators whose period of oscillation is independent of the energy or amplitude are called isochronous systems \cite{Calogero_book}. For rational potentials, the following two are the only ones that support isochronous oscillations throughout the accessible region of phase space \cite{Chalykh_2005}: 
\begin{eqnarray}
V_1(x) &=& \frac{ \omega^2 x^2}{2}, \quad \quad ~~~~~~x \in \mathbb{R}, \label{harm} \\
V_2(x) &=& Ax^2 + \frac{B}{x^2}, \quad \quad x \in \mathbb{R}_+, \label{iso}
\end{eqnarray} with $\omega, A, B > 0$. While $V_1(x)$ is the familiar harmonic potential, leading to the linear dynamics, $V_2(x)$ is called the isotonic potential \cite{Weissman_1979}, whose equation of motion is the Ermakov-Pinney equation \cite{Morris_2015}. If a nonlinear dynamical system cannot be derived from a constant-mass Hamiltonian with potential $V_1$ or $V_2$, a standard way to check for isochronicity is to seek a map to the harmonic-oscillator equation \cite{Urabe_1961,Guha_2013}, sometimes via a nonlocal transformation \cite{Guha_2019}.

\vspace{2mm}

A generic family of systems which is of much interest in the theory of dynamical systems is the family of Li\'enard equations
\begin{equation}\label{lienard1}
\ddot{x} + f(x) \dot{x} + g(x) = 0,
\end{equation} for appropriate functions $f(x)$ and $g(x)$. Such a system is generally not derivable from a standard Hamiltonian written as the sum of a kinetic energy and a potential. The linearization of the system (\ref{lienard1}) to the form $\ddot{X} + \omega^2 X = 0$ directly leads to the isochronicity condition \cite{Guha_2019}  (see \cite{Pandey_2009a,Pandey_2009b} for discussions on linearization to free-particle dynamics)
\begin{equation}\label{Cond}
  g(x) = \omega^2 x + \frac{1}{x^3} \bigg[ \int_0^x x' f(x') dx' \bigg]^2,
  \end{equation} and one particular case is $f(x) = kx$ and $g(x) = \omega^2 x + \frac{k^2}{9} x^3$, giving the system $\ddot{x} + kx \dot{x} + \omega^2 x + \frac{k^2}{9} x^3 = 0$, whose isochronicity is well known \cite{Chandrasekar_2005,GhoseChoudhury_2017}. The generic isochronicity condition (\ref{Cond}) was known from earlier treatments as well \cite{Sabatini_1999,Christopher_2004}. It may be mentioned that the equation $\ddot{x} + kx \dot{x} + \omega^2 x + \frac{k^2}{9} x^3 = 0$ is called the modified Emden equation \cite{Pandey_2009a,Pandey_2009b,Chandrasekar_2005a,Chandrasekar_2006,Lakshmanan_2013,Tiwari_2015,Mustafa_2023}.

\vspace{2mm}

A tempting conjecture for isochronous systems is that their quantum spectra are equispaced \cite{Carinena_2007}. This holds for the potentials (\ref{harm}) and (\ref{iso}), and for isochronous systems whose Schr\"odinger equations can be mapped to those with these potentials by a suitable change of variable \cite{GhoseChoudhury_2013,Bagchi_2015a,Bagchi_2025,Ghosh_2025} (see \cite{Carinena_2008,Quesne_2008,Levai_2023} for related potentials); however, the correspondence does not hold in general \cite{Dorignac_2005}. The aim of the present paper is to follow up on the recent development \cite{Bagchi_2025} to expose new insights on isochronous Li\'enard systems with a focus on branched Hamiltonians. We shall demonstrate that in contrast to the unbranched case yielding an exactly-equispaced spectrum \cite{Bagchi_2015a,Bagchi_2025}, the branched Hamiltonians produce quasi-harmonic level spacings that cluster around $2\omega$, along with small deviations. 

\section{Hamiltonian aspects}\label{Ham_sec}
Our aim is to study Li\'enard systems to assess their quantum spectra. The formal procedure of canonical quantization requires one to know the Hamiltonian of the system as a function of the canonical variables $x$ and $p$. While it is not immediately clear whether the system (\ref{lienard1}) can be derived from a Lagrangian or Hamiltonian function, it has been shown earlier \cite{GhoseChoudhury_2017,Nucci_2010} (see also, \cite{Bagchi_2015a,Bagchi_2025,Bagchi_2024}) that should $f(x)$ and $g(x)$ satisfy the so-called Chiellini condition
\begin{equation}\label{Chiellini_condition}
\frac{d}{dx} \bigg(\frac{g(x)}{f(x)}\bigg) + \ell (\ell + 1) f(x) = 0,
\end{equation} with $\ell$ being a real constant, an analytical form of the Lagrangian can be found. As a matter of fact, two distinct roots for $\ell$ can result in two distinct forms of Lagrangians, each of which can give rise to equation (\ref{lienard1}). One can thus find the corresponding Hamiltonian functions. Straightforward manipulations (see \cite{Bagchi_2015a,Bagchi_2025,Bagchi_2024} for details) reveal that the associated Lagrangians are
\begin{equation}
L(x,\dot{x})_\ell = \frac{\ell^2}{(\ell+1)(2\ell + 1)} \bigg(\dot{x} - \frac{g(x)}{\ell f(x)} \bigg)^{\frac{2\ell+1}{\ell}} \quad \implies \quad p = \frac{\partial L}{\partial \dot{x}} = \bigg(\frac{\ell}{\ell+1}\bigg) \bigg(\dot{x} - \frac{g(x)}{\ell f(x)} \bigg)^{\frac{\ell+1}{\ell}}.
\end{equation}
First, let us note that $\ell = 0, -1, -\frac{1}{2}$ are to be excluded. Second, let us also note that if $\frac{\ell}{\ell+1}$ is not an integer, solving for $\dot{x}$ as a function of $p$ will lead to roots or branches. The Hamiltonian functions are
\begin{equation}\label{Ham}
H_\ell = \frac{g(x)}{\ell f(x)} p + \frac{\ell}{2\ell+1} \bigg( \frac{\ell + 1}{\ell} p \bigg)^{\frac{2\ell + 1}{\ell + 1}},
\end{equation}
where $\{x,p\}=1$ for the canonical Poisson bracket. If branching is present due to non-integral $\frac{\ell}{\ell+1}$, it must also be taken into account. It is convenient to define a rescaled momentum $\tilde{p} = \frac{\ell + 1}{\ell}p$, allowing us to express the Hamiltonian (\ref{Ham}) as
\begin{equation}\label{Ham1}
H_\ell = \frac{g(x)}{(\ell + 1) f(x)} \tilde{p} + \frac{\ell}{2\ell+1} \tilde{p}^{\frac{2\ell + 1}{\ell + 1}},
\end{equation} where $\{x,\tilde{p}\} = \frac{\ell + 1}{\ell}$. 

\vspace{2mm}

\noindent
\textbf{Remark:} The system of our interest, which is $\ddot{x} + kx\dot{x} + \omega^2 x + \frac{k^2}{9} x^3 = 0$, is a special case of the Kukles family of equations
   \begin{equation}\label{Kukles}
   \ddot{x} + (\alpha_0 + \alpha_1 x + \alpha_2 x^2) \dot{x} + (\beta_1 x + \beta_2 x^2 + \beta_3 x^3) = 0,
   \end{equation} for real constants $\alpha_0, \alpha_1, \alpha_2, \beta_1, \beta_2$, and $\beta_3$. It satisfies the Chiellini condition (\ref{Chiellini_condition}) for appropriate choices of the underlying parameters. However, it can be shown (see Appendix (\ref{app})) that the only Li\'enard system with polynomial coefficients which is isochronous, i.e., satisfies the condition (\ref{Cond}) and also satisfies the Chiellini condition (\ref{Chiellini_condition}) (that is, admits a Hamiltonian description) is the modified Emden equation $\ddot{x} + kx\dot{x} + \omega^2 x + \frac{k^2}{9} x^3 = 0$ \cite{GhoseChoudhury_2017}, whose quantum aspects have been demonstrated in \cite{Bagchi_2015a,Bagchi_2025} (see also, the preceding work \cite{Ruby_2012}). No other polynomial choices for the functions $f(x)$ and $g(x)$ can satisfy the conditions (\ref{Cond}) and (\ref{Chiellini_condition}) simultaneously if $\omega > 0$. 

\section{Canonical quantization of the branched Hamiltonians}\label{can_sec}
The nonlinear Li\'enard system $\ddot{x} + k x \dot{x} + \omega^2 x + \frac{k^2}{9} x^3 = 0$ satisfies the isochronicity condition (\ref{Cond}) and also admits the Chiellini condition (\ref{Chiellini_condition}) for $\ell = -1/3, -2/3$, leading to analytical forms of the Hamiltonian functions \cite{GhoseChoudhury_2017,Bagchi_2015a,Bagchi_2025,Bagchi_2024}. For both $\ell = -1/3$ and $\ell = -2/3$, the two classes of Hamiltonians are expressible in the generic form \cite{Bagchi_2015a,Bagchi_2025,Bagchi_2024,Ruby_2012,Bagchi_2015b,Bagchi_2019}
\begin{equation}
H = \frac{x^2}{2m(p)} + V(p),
\end{equation} for appropriate functions $m(p)$ and $V(p)$. Thus it appears as if the roles of position and momentum have become interchanged, accompanied by the appearance of a momentum-dependent mass. The formal procedure of canonical quantization requires careful ordering of the kinetic-energy-like operator due to the fact that position and momentum do not commute in quantum mechanics. To this end, let us apply the ordering strategy due to von Roos \cite{vonRoos_1983}, giving the general form of a Hermitian ordering as
\begin{equation}
\hat{T} = \frac{1}{4} \left[ m^\alpha(\hat{p}) \hat{x} m^\beta(\hat{p}) \hat{x} m^\gamma(\hat{p}) + m^\gamma(\hat{p}) \hat{x} m^\beta(\hat{p}) \hat{x} m^\alpha(\hat{p}) \right],
\end{equation}
where $(\alpha, \beta, \gamma)$ are the ambiguity parameters satisfying the constraint $\alpha + \beta + \gamma = -1$. Using the canonical quantization (we will take $\hbar = 1$): $\hat{p} = p$ and $\hat{x} = i \frac{d}{dp}$, the kinetic-energy-like operator acting on a time-independent wavefunction as $\hat{T}\psi(p)$ is given by \cite{Bagchi_2015a,Bagchi_2025}
\begin{equation}
\hat{T} = - \frac{1}{2 m(p)} \bigg[\frac{d^2}{dp^2} - \frac{m'(p)}{m(p)} \frac{d}{dp} + \frac{\beta + 1}{2} \bigg( 2 \frac{m'(p)^2}{m(p)^2} - \frac{m''(p)}{m(p)}\bigg) + \alpha(\alpha + \beta + 1) \frac{m'(p)^2}{m(p)^2} \bigg],
\label{vonroosT}
\end{equation} where the primes denote differentiation with respect to $p$. There are two different forms of $m(p)$ and $V(p)$ that can describe the system under consideration \cite{Bagchi_2025}. One of them (that for $\ell=-2/3$) admits a singular momentum-dependent-potential profile whose time-independent Schr\"odinger equation can be mapped to that of a particle with constant mass in the isotonic potential (\ref{iso}) \cite{Bagchi_2015a,Bagchi_2025}. As a result, the spectrum is equispaced and the eigenfunctions are expressible using associated Laguerre polynomials \cite{Weissman_1979}. The other form of the Hamiltonian (that for $\ell=-1/3$) is the branched pair \cite{Bagchi_2015a,Bagchi_2025,Bagchi_2024}
\begin{equation}\label{H13}
H^\pm = -3 p \bigg( \frac{kx^2}{9} + \frac{\omega^2}{k} \bigg) \mp \sqrt{-2p}, \quad \quad \quad p < 0,
\end{equation}
where the two signs before the square root originate from the generalized velocity being a multivalued function of $p$ \cite{Bagchi_2024}. Defining $\tilde{p} = -2p > 0$ and using the expression (\ref{vonroosT}), the time-independent Schr\"odinger equation $(\hat{T} + V(\hat{p})) \psi = E\psi$  is given by
\begin{equation}
\tilde{p} \frac{d^2 \psi_\pm(\tilde{p})}{d\tilde{p}^2} + \frac{d \psi_\pm(\tilde{p})}{d\tilde{p}} + \frac{\alpha(\alpha + \beta + 1)}{\tilde{p}} \psi_\pm(\tilde{p}) + \frac{3}{2k} (E - V^\pm(\tilde{p})) \psi_\pm(\tilde{p}) = 0. 
\end{equation} 
Let us now assess this further. 

\subsection{Possibility of a polynomial solution}
Performing a change of variable as $\tilde{p} = (k/24) \xi^2$ and writing $\phi_\pm(\xi) = \sqrt{\xi} \psi_\pm (\xi)$, we can transform the above-mentioned equation to the following form \cite{Bagchi_2025}:
 \begin{equation}\label{phipmeqn}
- \frac{d^2 \phi_\pm(\xi)}{d\xi^2} + \bigg[ \frac{\omega^2 (\xi \mp \xi_0)^2}{64} + \frac{\epsilon - \frac{1}{4}}{\xi^2} - \frac{k}{24 \omega^2} \bigg] \phi_\pm(\xi) =  \frac{E}{4} \phi_\pm(\xi), \quad \quad \xi_0 = \sqrt{\frac{k}{6}} \frac{4}{\omega^2},
\end{equation} with $\xi > 0$ and $\epsilon = - 4 \alpha(\alpha + \beta +1) = 4 \alpha \gamma$. Thus one gets the effective potentials 
\begin{equation}\label{effpotpm}
V_{\rm eff}^\pm(\xi) = \frac{\omega^2 (\xi \mp \xi_0)^2}{64} + \frac{\epsilon - \frac{1}{4}}{\xi^2} - \frac{k}{24 \omega^2}, \quad \quad \xi >0,
\end{equation} which resemble the isotonic potential but with shifted centers when $\xi_0 \neq 0$. The special case $\epsilon = \frac{1}{4}$ that arises for $\alpha = \gamma = -\frac{1}{4}$ and $\beta = -\frac{1}{2}$ was studied in the earlier work \cite{Bagchi_2025}, in which case the centrifugal term vanishes, leading to some simplification. The corresponding effective potentials that read 
\begin{equation}
V_{\rm eff}^\pm(\xi) = \frac{\omega^2 (\xi \mp \xi_0)^2}{64} - \frac{k}{24 \omega^2}, \quad \quad \xi >0,
\end{equation} resemble the harmonic potential but with the boundary condition that the wavefunctions must vanish at $\xi = 0$ (and additionally at $\xi \to \infty$ as required for normalizability). This condition severely restricts the possibility of obtaining polynomial solutions in terms of the Hermite polynomials. However, as shown in \cite{Bagchi_2025}, if the following condition holds:
\begin{equation}\label{hermite_cond}
H_n \bigg( \sqrt{\frac{\omega \xi_0^2}{8} }\bigg) = 0,
\end{equation} for some $n \in \mathbb{N}$, then 
\begin{equation}
(\phi_\pm)_n(\xi) \sim e^{- \frac{\omega}{16}(\xi \mp \xi_0)^2} H_n \left( \sqrt{\frac{\omega}{8}} (\xi \mp \xi_0) \right)
\end{equation} is one bound-state solution. Since the zeros of Hermite polynomials are simple and do not repeat across $n$, the above-mentioned condition can be satisfied at most by only one value of $n$, meaning that there is at most one bound state expressible via a Hermite polynomial.

\section{Non-polynomial solutions and energy quantization}\label{num_sec}
Let us now move beyond the existing results to assess the quantum mechanics dictated by the branched Hamiltonians $H_\pm$ that lead to the effective potentials (\ref{effpotpm}). While for the case $\epsilon =\frac{1}{4}$, the possibility of polynomial truncation was studied in \cite{Bagchi_2025} and pointed out in the previous section, we will now revisit the eigenproblem (\ref{phipmeqn}) subject to the Dirichlet boundary condition $\phi_\pm(0) = 0$, along with $\lim_{\xi \to \infty} \phi_\pm (\xi) = 0$. In what follows, we shall take $k$ to be small, where the extent of smallness will be quantified in Sec. (\ref{pert_sec}).

\subsection{$\epsilon = \frac{1}{4}$}
\begin{figure}[h]
    \centering
    \includegraphics[width=0.7\linewidth]{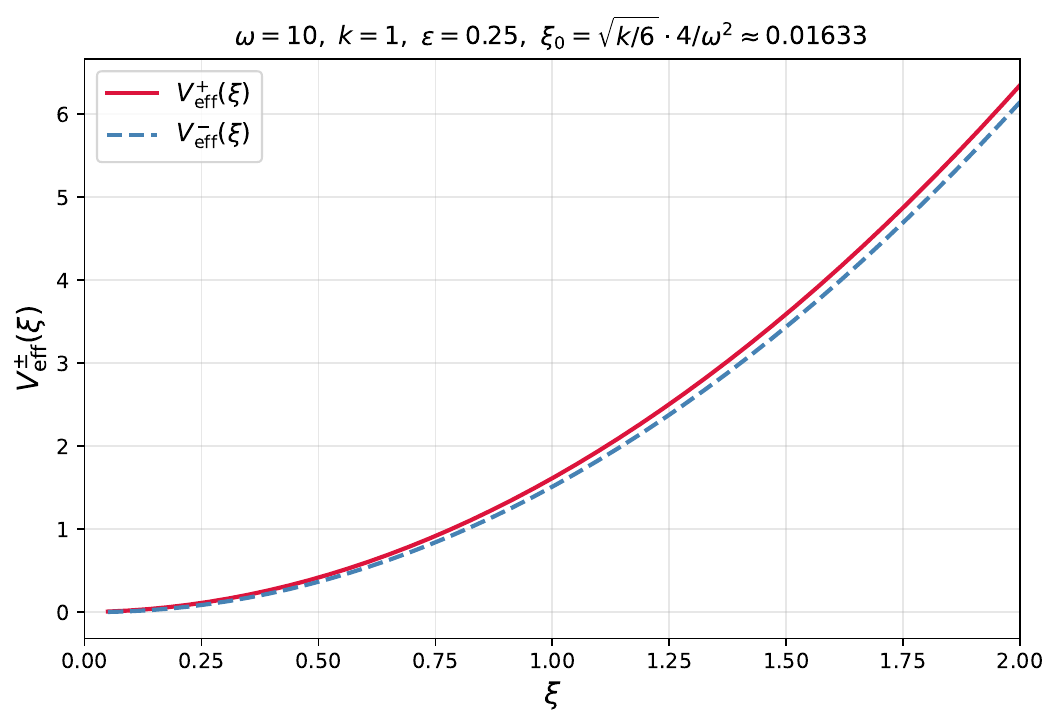}
    \caption{Effective potentials $V^\pm_{\rm eff}(\xi)$ for $\epsilon = \frac{1}{4}$.}
    \label{fig1}
\end{figure}

For $\epsilon=\frac{1}{4}$, the eigenproblem (\ref{phipmeqn}) reduces to 
\begin{equation}
-\frac{d^2\phi_\pm(\xi)}{d\xi^2}
+ \frac{\omega^2(\xi\mp\xi_0)^2}{64} \phi_\pm(\xi)
=\lambda \phi_\pm(\xi),
\qquad 
\lambda=\frac{E}{4}+\frac{k}{24\omega^2}.
\label{non_singular}
\end{equation}
Fig. (\ref{fig1}) shows the corresponding effective potentials. 
The general square-integrable solution of the displaced oscillator is given by the parabolic cylinder functions \cite{Abramowitz_Stegun_1965} as
\begin{equation}
\phi_\pm(\xi)=D_\mu\left(\sqrt{\frac{\omega}{4}}(\xi\mp\xi_0)\right),
\qquad
\mu=\frac{E}{\omega}+\frac{k}{6\omega^3}-\frac{1}{2}.
\label{parabolic_cylinder}
\end{equation}
The boundary condition $\phi_\pm(0)=0$ enforces following the transcendental quantization:
\begin{equation}
D_\mu\left(\mp\sqrt{\frac{\omega}{4}}\xi_0\right)=0, 
\qquad 
E_n=\omega\left(\mu_n+\frac{1}{2}\right)-\frac{k}{6\omega^2}.
\label{quantization_condition}
\end{equation}
Unlike the full-line harmonic oscillator admitting solutions with both even and odd parity, or the half-line harmonic oscillator admitting only solutions with odd parity (both cases admit equispaced spectra), here the fact that the center at $\xi_0$ does not coincide with the boundary at $\xi = 0$ destroys parity symmetry. The spectrum can be evaluated by solving equation (\ref{non_singular}) numerically and the first six eigenvalues are shown in Table (\ref{table1}). These energies agree within numerical tolerance with those obtained by numerically solving the transcendental condition (\ref{quantization_condition}). The spectrum is approximately equispaced with a spectral spacing $\simeq 2\omega$. 
\begin{table}[h!]
\centering
\caption{First six eigenvalues from direct numerical solution for $\omega = 10$, $k=1$, and $\epsilon = \frac{1}{4}$.}
\begin{tabular}{l c r}
\toprule
$n$ & $E^+_n$ & $E^-_n$ \\
\midrule
0 & 14.77286544 & 15.18436212 \\
1 & 34.65936779 & 35.27686069 \\
2 & 54.57419677 & 55.34619560 \\
3 & 74.50319290 & 75.40394802 \\
4 & 94.44103893 & 95.45445609 \\
5 & 114.38507520 & 115.49988822 \\
\bottomrule
\end{tabular}
\label{table1}
\end{table}

\subsection{$\epsilon\neq\frac{1}{4}$}
\begin{figure}[h]
    \centering
    \includegraphics[width=0.7\linewidth]{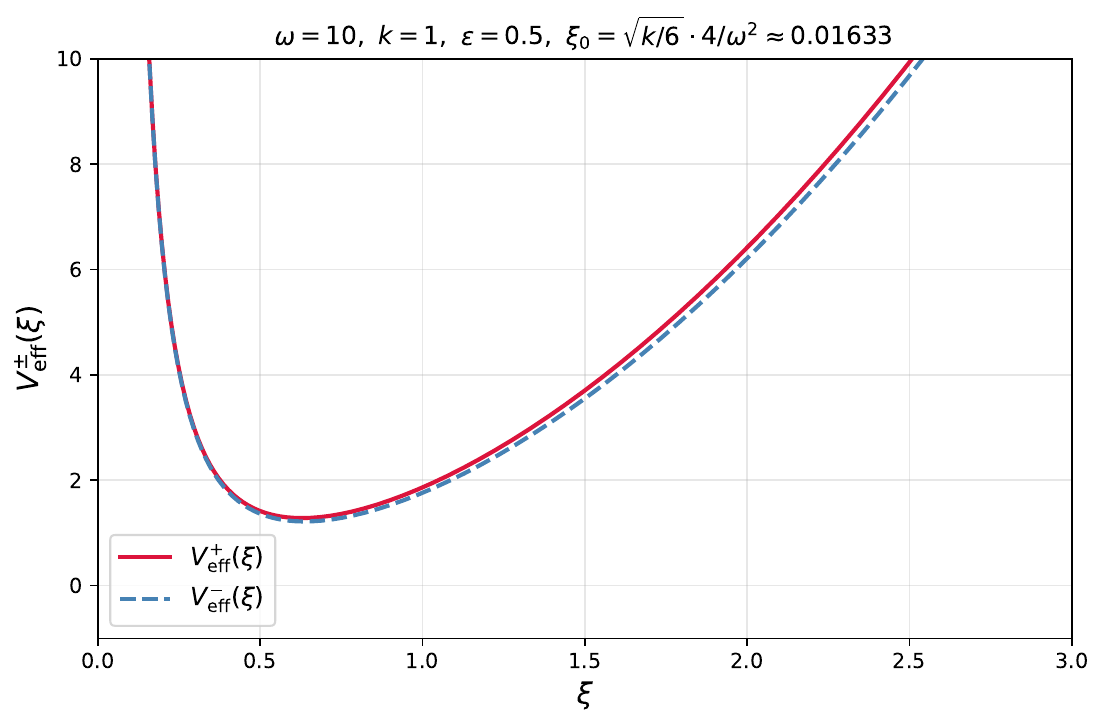}
    \caption{Effective potentials $V^\pm_{\rm eff}(\xi)$ for $\epsilon = \frac{1}{2}$.}
    \label{fig2}
\end{figure}

\begin{figure}[h]
    \centering
    \includegraphics[width=0.7\linewidth]{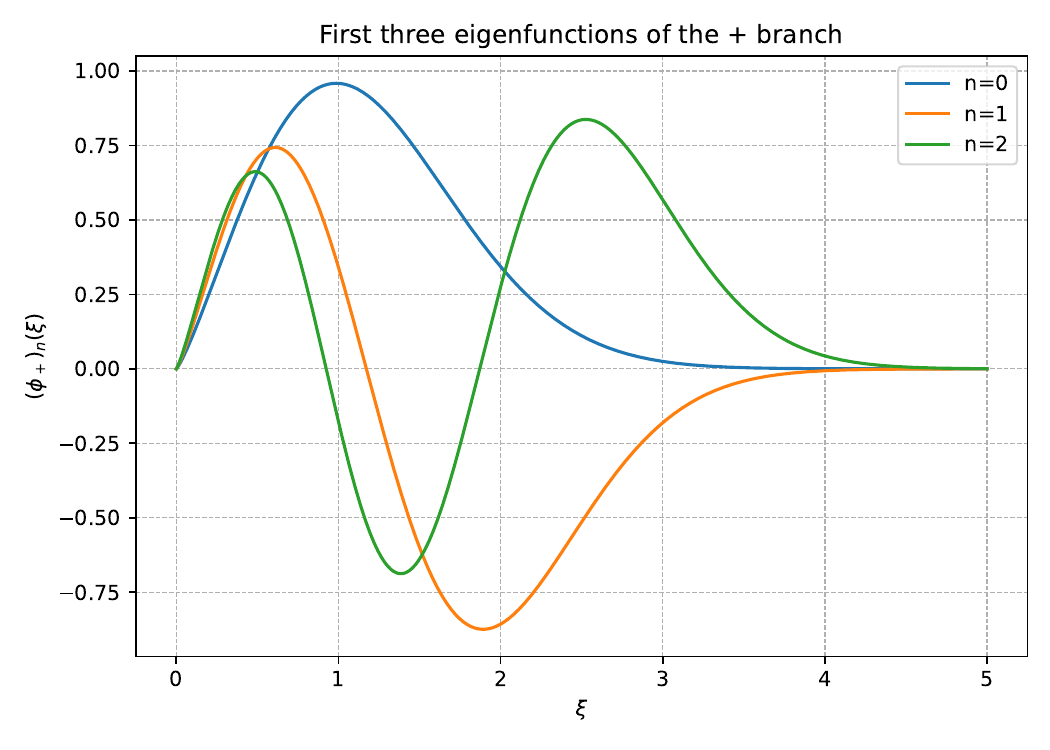}
    \caption{Eigenfunctions $(\phi_+)_n(\xi)$ for $n=0,1,2$, taking $\omega = 10$, $k=1$, and $\epsilon = \frac{1}{2}$. The corresponding eigenvalues read $E^+_0 \simeq 16.85$, $E^+_1 \simeq 36.75$, and $E^+_2 \simeq 56.68$.}
    \label{eigenfuncs_plusbranch_0to5}
\end{figure}

\begin{figure}[h]
    \centering
    \includegraphics[width=0.7\linewidth]{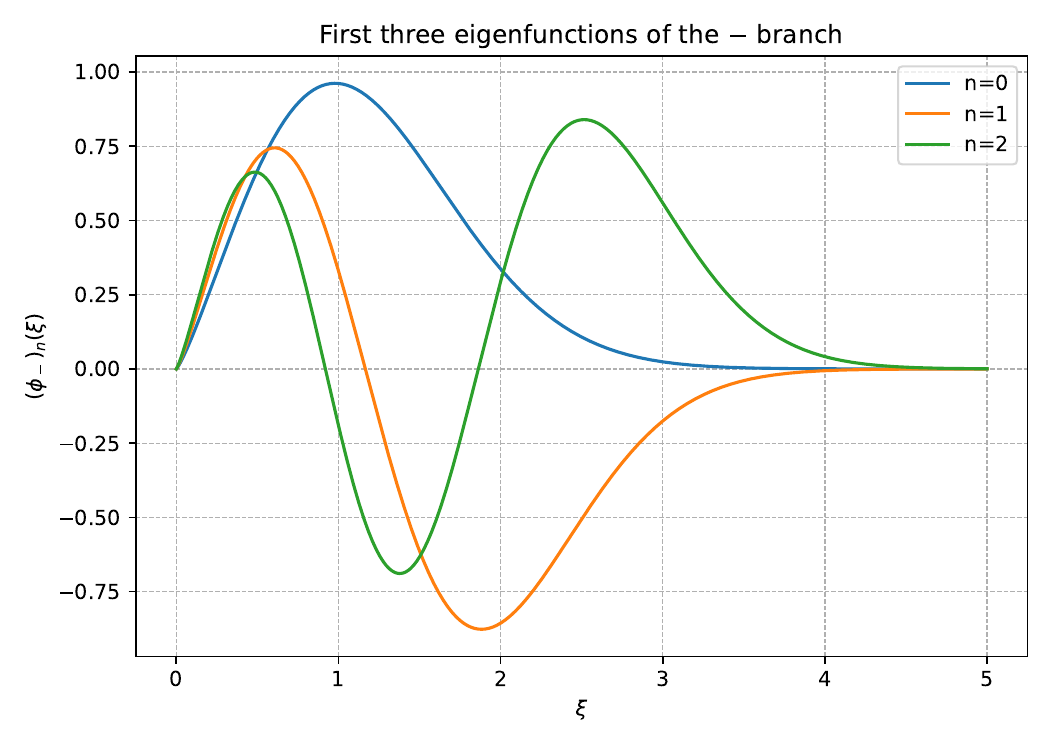}
    \caption{Eigenfunctions $(\phi_-)_n(\xi)$ for $n=0,1,2$, taking $\omega = 10$, $k=1$, and $\epsilon = \frac{1}{2}$. The corresponding eigenvalues read $E^-_0 \simeq 17.29$, $E^-_1 \simeq 37.39$, and $E^-_2 \simeq 57.46$.}
    \label{eigenfuncs_minusbranch_0to5}
\end{figure}

With $\epsilon\neq\frac{1}{4}$, the inverse-square term remains non-zero and equation (\ref{phipmeqn}) can be expressed as
\begin{equation}
-\frac{d^2\phi_\pm(\xi)}{d\xi^2}
+\left[
\frac{\omega^2(\xi\mp\xi_0)^2}{64}
+\frac{\epsilon-\frac{1}{4}}{\xi^2}
\right]\phi_\pm(\xi)
=\lambda \phi_\pm(\xi),
\qquad 
\lambda=\frac{E}{4}+\frac{k}{24\omega^2}.
\label{singular_eq}
\end{equation}
Fig. (\ref{fig2}) shows the corresponding effective potentials. 
For $\xi_0\neq0$, the above equation cannot be exactly reduced to a confluent-hypergeometric form because the singular term $\xi^{-2}$ and the displaced harmonic term $(\xi\mp\xi_0)^2$ have different centers. The solution must therefore be obtained numerically, subject to the physical boundary conditions
\begin{equation}
\phi_\pm(\xi)\sim \xi^{\frac{1}{2}+\sqrt{\epsilon}} \quad (\xi\to0), 
\qquad 
\phi_\pm(\xi)\sim e^{-\frac{\omega}{16}(\xi\mp\xi_0)^2}\quad (\xi\to\infty).
\end{equation}
The first three eigenfunctions (up to an overall sign and normalization) for the two branches ($+$ and $-$) have been displayed in Figs. (\ref{eigenfuncs_plusbranch_0to5}) and (\ref{eigenfuncs_minusbranch_0to5}). 
The spectrum can be evaluated numerically and the first six eigenvalues are shown in Table (\ref{table2}). 
\begin{table}[h!]
\centering
\caption{First six eigenvalues from direct numerical solution for $\omega = 10$, $k=1$, and $\epsilon = \frac{1}{2}$.}
\begin{tabular}{l c r}
\toprule
$n$ & $E^+_n$ & $E^-_n$ \\
\midrule
0 & 16.84868053 & 17.29250939 \\
1 & 36.75113362 & 37.38995201 \\
2 & 56.67606076 & 57.46485124 \\
3 & 76.61285412 & 77.52785010 \\
4 & 96.55723877 & 97.58322724 \\
5 & 116.50700377 & 117.63319363 \\
\bottomrule
\end{tabular}
\label{table2}
\end{table}
The levels are thus approximately equispaced with the spacing $\simeq 2\omega$. It is worth emphasizing that for general $\epsilon$, the wavefunctions can be expressed using a generalized Frobenius series which does not generally truncate and the Dirichlet boundary condition $\phi_\pm(0) = 0$ then leads to a transcendental quantization condition whose numerical solution gives the energies listed in Tables (\ref{table1}) and (\ref{table2}) within numerical tolerance. Based on the numerical spectrum for the parameters tested, a linear fit can be suggested in the form
\begin{equation}
E^\pm_n \simeq \omega(2n + \delta_\pm),
\end{equation} for constants $\delta_+$ and $\delta_-$. This is a good estimate of the low-lying spectrum.

\section{Perturbation theory for small anharmonic corrections}\label{pert_sec}
In this section, we will treat the anharmonicity parameter $k$ perturbatively. To this end, let us note that the eigenproblem (\ref{phipmeqn}) can be expressed as
\begin{equation}
-\frac{d^2\phi_\pm}{d\xi^2}
+\left[\frac{\omega^2 \xi^2}{64}+\frac{\epsilon-\frac{1}{4}}{\xi^2} \mp \sqrt{\frac{k}{6}} \frac{\xi}{8} \right]\phi_\pm
=\frac{E}{4}\phi_\pm,
\label{main}
\end{equation} subject to the Dirichlet boundary condition $\phi_\pm(0)=0$, and $\lim_{\xi \rightarrow \infty} \phi_\pm(\xi) = 0$. For $k=0$, the problem corresponds to the isotonic oscillator and can be exactly solved, leading to an equispaced spectrum. For small $k$, therefore, one can treat $k$ perturbatively to quantify anharmonic corrections that make the spectrum quasi-harmonic. For perturbative control of the eigenproblem (\ref{main}), the linear term $\mp \sqrt{k/6}(\xi/8)$ must remain much smaller than both the quadratic confinement and the unperturbed level spacing. In order of magnitude, this requires $k \ll \frac{3\omega^{3}}{n_{\rm max}+\sqrt{\epsilon}}$ if $n_{\rm max}$ is the largest quantum number that one cares about. To begin, let us note that when $k=0$, branching does not apply and the exact solutions are \cite{Weissman_1979}
\begin{eqnarray}
\phi_n(\xi) &=& N_n \xi^{\frac{1}{2} + \sqrt{\epsilon}} e^{-\frac{\omega \xi^2}{16}} L^{\sqrt{\epsilon}}_n \bigg( \frac{\omega \xi^2}{8} \bigg), \quad \quad N_n = \sqrt{\frac{2(\omega/8)^{\sqrt{\epsilon}+1}n!}{\Gamma(n+\sqrt{\epsilon}+1)}}, \\
E^{(0)}_n &=& \omega(2n + \sqrt{\epsilon} + 1).
\end{eqnarray}
The orthogonality of these solutions directly follows from that of the associated Laguerre polynomials \cite{Abramowitz_Stegun_1965}. One immediately notes a spectral spacing of $2\omega$. It may be noted that the $k=0$ limit is singular in the classical Hamiltonian (\ref{H13}) but can be taken in the Schr\"odinger equation (\ref{main}). Employing the standard time-independent perturbation theory \cite{Bender_Orszag_1999}, the first-order correction ($\sim \mathcal{O}(\sqrt{k})$) to the spectrum is given by
\begin{equation}
(\Delta E)^\pm_n = \mp \sqrt{\frac{k}{24}} \int_{0}^{\infty}\xi \phi_{n}(\xi)^2 d\xi. 
\end{equation}
Using the analytical form of $\phi_n(\xi)$ quoted above, we can express the integral as
\begin{equation}\label{master}
 \int_{0}^{\infty}\xi \phi_{n}(\xi)^2 d\xi = \left( \frac{\omega}{8} \right)^{-\frac{1}{2}} \frac{n!}{\Gamma(n+\sqrt{\epsilon}+1)}
\int_0^\infty t^{\sqrt{\epsilon}+\frac{1}{2}} e^{-t}L_n^{\sqrt{\epsilon}}(t)^2 dt,
\end{equation}
where $t = (\omega/8) \xi^2$. Thus, using the hypergeometric expansion \cite{Abramowitz_Stegun_1965}
\begin{equation}
L_n^{\sqrt{\epsilon}}(t)=\sum_{j=0}^{n}\frac{(-n)_j}{(\sqrt{\epsilon}+1)_j} \frac{t^j}{j!},
\label{Lag_series}
\end{equation} where $(a)_j$ is the Pochhammer symbol, integrating term-by-term gives
\begin{align}
\int_0^\infty t^{\sqrt{\epsilon}+\frac{1}{2}} e^{-t} L_n^{\sqrt{\epsilon}}(t)^2 dt
&=\sum_{j=0}^{n}\sum_{l=0}^{n}
\frac{(-n)_j (-n)_l}{(\sqrt{\epsilon}+1)_j (\sqrt{\epsilon}+1)_l}
\frac{1}{j! l!}
\int_0^\infty t^{\sqrt{\epsilon}+\frac{1}{2}+j+l} e^{-t} dt \nonumber
\\
&=\sum_{j=0}^{n}\sum_{l=0}^{n}
\frac{(-n)_j (-n)_l}{(\sqrt{\epsilon}+1)_j (\sqrt{\epsilon}+1)_l}
\frac{\Gamma\bigl(\sqrt{\epsilon}+\frac{3}{2}+j+l\bigr)}{j! l!}.
\label{double_sum}
\end{align}
Substituting the result (\ref{double_sum}) into the expression (\ref{master}) gives the final form which goes as
\begin{equation}
 \int_{0}^{\infty}\xi \phi_{n}(\xi)^2 d\xi 
=\bigg(\frac{\omega}{8}\bigg)^{-\frac{1}{2}} \frac{n!}{\Gamma(n+\sqrt{\epsilon}+1)}
\sum_{j=0}^{n}\sum_{l=0}^{n}
\frac{(-n)_j (-n)_l}{(\sqrt{\epsilon}+1)_j (\sqrt{\epsilon}+1)_l}
\frac{\Gamma\bigl(\sqrt{\epsilon}+\frac{3}{2}+j+l\bigr)}{j! l!}. 
\end{equation}
Therefore we have now found the corrected energies of the branched partners as given by
\begin{equation}
E^{\pm, (1)}_n = \omega(2n + \sqrt{\epsilon} + 1) \mp \sqrt{\frac{k}{24}} \bigg(\frac{\omega}{8}\bigg)^{-\frac{1}{2}} \frac{n!}{\Gamma(n+\sqrt{\epsilon}+1)}
\sum_{j=0}^{n}\sum_{l=0}^{n}
\frac{(-n)_j (-n)_l}{(\sqrt{\epsilon}+1)_j (\sqrt{\epsilon}+1)_l}
\frac{\Gamma\bigl(\sqrt{\epsilon}+\frac{3}{2}+j+l\bigr)}{j! l!}. 
\end{equation}
Taking $\omega = 10$, $k=1$, and $\epsilon = \frac{1}{2}$, the first six eigenvalues are displayed in Table (\ref{table3}) and are in good agreement with those presented in Table (\ref{table2}). This demonstrates that perturbation theory up to $\mathcal{O}(\sqrt{k})$ is quite accurate for such parameter choices in describing the low-lying spectrum.
\begin{table}[h!]
\centering
\caption{First six eigenvalues from (analytic) perturbation theory up to $\mathcal{O}(\sqrt{k})$ with the parameter values $\omega = 10$, $k=1$, and $\epsilon = \frac{1}{2}$.}
\begin{tabular}{l c r}
\toprule
$n$ & $E^{+,(1)}_n$ & $E^{-,(1)}_n$ \\
\midrule
0 & 16.856 & 17.286 \\
1 & 36.750 & 37.392 \\
2 & 56.668 & 57.474 \\
3 & 76.603 & 77.539 \\
4 & 96.549 & 97.593 \\
5 & 116.503 & 117.639 \\
\bottomrule
\end{tabular}
\label{table3}
\end{table}

\section{Discussion}\label{disc_sec}
The modified Emden equation $\ddot{x} + kx\dot{x} + \omega^2 x + \frac{k^2}{9} x^3 = 0$, quantized via the von Roos prescription, yields effective half-line oscillators, and a particular form of the Hamiltonian emphasized in \cite{Bagchi_2015a,Bagchi_2025} leads to exact solutions in terms of the associated Laguerre polynomials, accompanied by an equispaced spectrum. However, considering the branched Hamiltonians (\ref{H13}) which have been the main focus of this paper, one encounters (when $k$ is small enough to be regarded as a perturbation for the first few levels) only approximate harmonic spacings between the levels -- quasi-harmonic -- with the spacings clustering around $2\omega$, along with small deviations. In the case $\epsilon \neq \frac{1}{4}$, the near-origin index $s = \frac{1}{2} + \sqrt{\epsilon}$ ensures regularity, while the shifted quadratic confinement prevents an exact confluent-hypergeometric reduction, accounting for the mild anharmonicity observed numerically; for $\epsilon = \frac{1}{4}$, the spectrum follows parabolic-cylinder quantization. These results clarify that isochronous classical behavior does not necessarily imply perfectly-harmonic quantum spectra. \\

\noindent
\textbf{Acknowledgements:} A.G. thanks Akash Sinha for related discussions and also thanks the Ministry of Education, Government of India for financial support in the form of a Prime Minister’s Research Fellowship (ID: 1200454) during the initial stages of this study. A.G.C. and P.G. gratefully acknowledge discussions with Pepin Cari\~nena. 

\appendix 
\counterwithin*{equation}{section}
\renewcommand\theequation{\thesection\arabic{equation}}

\section{Uniqueness of the system satisfying the conditions (\ref{Cond}) and (\ref{Chiellini_condition}) when $f(x)$ and $g(x)$ are polynomials}\label{app}
Let us consider situations in which $f(x)$ is a polynomial. It is easy to check that when $f(x)$ is a real constant, i.e., a monomial of degree zero, the conditions (\ref{Cond}) and (\ref{Chiellini_condition}) are simultaneously satisfied for complex-valued $\ell$, a case that we shall exclude. As for non-constant but polynomial $f(x)$, the following is true: 
\begin{prop}
Let $f(x)$ be a real polynomial and $I(x)=\int_0^x x' f(x') dx'$. Then the isochronicity condition (\ref{Cond}) which is equivalent to 
\begin{equation}\label{iso_I}
g(x)=\omega^2 x+\frac{I(x)^2}{x^3},
\end{equation}
with fixed $\omega > 0$, is compatible with the Chiellini condition (\ref{Chiellini_condition}) if and only if $f(x)=kx$. In particular, no
polynomial of degree $\ge 2$ and no affine linear $kx+b$ with $b\ne 0$ works.
\end{prop}

\textbf{Proof --} Let us begin by noting that the Chiellini condition (\ref{Chiellini_condition}) is equivalent to the following identity:
\begin{equation}\label{poly-id}
\frac{dg(x)}{dx} f(x)-g(x) \frac{df(x)}{dx}+\ell(\ell+1)f(x)^3= 0.
\end{equation}

\noindent\textit{(1) Exclusion of degrees ${n\geq 2}$:}
Let $f(x)=a_nx^n+\cdots$ with $n\ge2$. Then the condition (\ref{iso_I}) gives
\begin{equation}
I(x)=\frac{a_n}{n+2}x^{n+2}+\cdots,\quad \quad
g(x)=\omega^2x+\frac{a_n^2}{(n+2)^2}x^{2n+1}+\cdots.
\end{equation}
This form of $g(x)$ does not satisfy the condition (\ref{poly-id}), as can be easily checked by substitution. 

\smallskip

\noindent\textit{(2) Feasible choice of ${f(x)}$:} Let $f(x)=kx+b$, so that
\begin{equation}
I(x)=\frac{k}{3}x^3+\frac{b}{2}x^2,\quad \quad
g(x)=\frac{k^2}{9}x^3+\frac{kb}{3}x^2+\Big(\omega^2+\frac{b^2}{4}\Big)x.
\end{equation}
Substituting into (\ref{poly-id}) and equating coefficients yields $\ell(\ell+1)=-2/9$, and
\begin{equation}
b\left(\frac{b^2}{36}+\omega^2\right)=0,
\end{equation}
which forces $b=0$.

\end{document}